\documentclass[mathleft
]{an}
\usepackage{graphicx}
\usepackage{times}
\begin{document}

% The following seven commands are intended for editorial usage and should be ignored by 
% the author(s).
\Pagespan{000}{000}% Document's page range. 
% If second parameter is left empty, the last page is computed automatically.
\Yearpublication{2009}%
\Yearsubmission{2008}%
\Month{0}%   
\Volume{0}%  
\Issue{0}% 
% \DOI{This.is/not.aDOI}% 

\title{Transverse motions in CSOs?}

\author{C. Stanghellini\inst{1}\fnmsep\thanks{Corresponding author:
  \email{cstan@ira.inaf.it}\newline}
%Example 
%for footnote, note the usage of the \texttt{fnmsep}
%command as separator between institute number and footnote mark} 
\and  D. Dallacasa\inst{1,2}
\and  T. Venturi \inst{1}
\and  T. An \inst{3}
\and  X.Y. Hong \inst{3}
}
\titlerunning{Transverse motions in CSOs?}
\authorrunning{Stanghellini et al.}
\institute{
Istituto di Radioastronomia -- INAF, Bologna, Italy
\and 
Dipartimento di Astronomia - Universit\`a di Bologna, Italy
\and 
Shanghai Observatory}

\received{08 Dec 2008}
\accepted{18 Dec 2008}
\publonline{later}

\keywords{Galaxies:active -- radio continuum:galaxies -- quasars:general}

\abstract{%
The measurement of proper motions in CSOs is a powerful tool to determine the dynamical evolution of the newly born extragalactic radio sources. 
We observed 3 CSOs with the VLBA in 2004 and in 2006 to monitor changes in their structure and measure the separation velocity of the hot spots. 
It is important to increase the size of the samples of CSOs with measured expansion velocity 
to test the existance of frustrated objects, and put stringent constraints on the current models.
We found for all the three objects observed a transverse motion of the hotspots, and we suggest as the more likey explanation 
a precession in the jet axis. This behaviour likely inhibits or at least slows down the radio source growth because the
head of the hotspot continuously hits new regions of the ISM. Therefore these radio sources may represent an old population of GPS/CSOs.
}

\maketitle

\section{Introduction}

Compact Symmetric Objects (CSOs) are radio sources with sub-kpc size  
with a morphology and radio luminosity similar to the classical doubles,
which are typically 1000 times larger. The integrated radio spectrum is generally
convex, with the peak frequency
around a few GHz. The projected linear size generally never
exceeds 1 kpc, and some sources are a few tens of pc in size only.

A key issue in the study of these sources is their age. 
They could be rather young ($<10^4$ yr) or substantially older ($>10^6$ yr).

It has also been suggested that many CSOs are short lived objects who
die before reaching large sizes (Readhead et al. 1994). 	

At present  there is large support to the youth
scenario. Expansion speeds detected are in agreement with 
estimated radiative ages to give to CSOs an age of a thousand to 
ten thousand years (Murgia et al. 1999, Polatidis and Conway 2003, Gugliucci et al. 2005).
Still there are the sources with upper limits
to their expansion speed,
and it is not demonstrated that all the sources are growing fast. 
In this context monitoring new CSOs increases the 
number of measurments of proper motions and improves the statistics.

\section{The selection of new target CSOs}

We searched CSOs
sources fulfilling the following requirements:

\noindent
$\bullet$ Simple double or triple structure on mas scale;

\noindent
$\bullet$ High declination, allowing a better sampling of the UV plane;

\noindent
$\bullet$ High flux density at 8.4 GHz, where the VLBA has the best combination
of sensitivity and resolution;

\noindent
$\bullet$ Size between 10 and $\sim$100 mas, to fully exploit the
spatial frequencies sampled by the VLBA.

\noindent
$\bullet$ Not already studied or under study for this purpose.

\begin{table}[!t]
\caption{Information on the observed radio sources }
\begin{center}
\begin{tabular}{ccccc}
\hline
\hline
source    & m$_r$   & z     &RA(J2000)&DEC(J2000)  \\
          &     &       &         &            \\
\hline
J0706+4647&   -  &    -  & 07 06 48.147&+46 47 56.21 \\
J1335+5844&  22.0   & 0.57  & 13 35 25.928&+58 44 00.29 \\
J1823+7938&   16.7& 0.224 & 18 23 14.109&+79 38 49.00 \\
\hline
\end{tabular}
\end{center}
\end{table}

Not many sources satisfied all these requirements.
A search in the literature and among the VLBI databases available
online was carried out
to select suitable sources belonging to the samples of GPS
radio sources and CSOs ( Snellen et al. 1998,
Stanghellini et al. 1998, Dallacasa et al. 2000,
Dallacasa et al. 2002a).
From the candidate sources we chose as the best suitable, the 3 objects listed
in table 1, which satisfy the above requirements.
Two objects do not have a measured redshift, but they
belong to GPS samples with planned/ongoing optical studies (e.g.
Dallacasa et al. 2002b), and in any case a rough redshift can be estimated for one of them (J1335+5844) by its optical magnitude as it is known that optical magnitudes of GPS/CSOs
correlate well with redshift (O'Dea et al. 1996, Snellen et al. 1996).

\begin{figure*}
      \includegraphics[angle=-90,width=13.5cm]{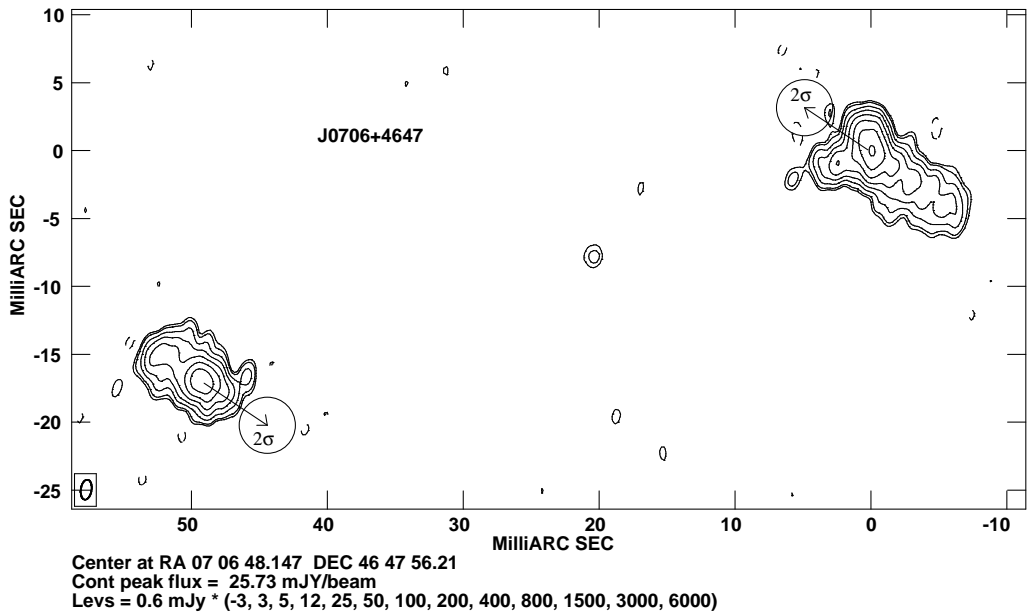}
\caption[]{J0706+4647, the arrows show the exprapolated motions in about 500 yr. See text for an explanation of the arrows. Here and in the next figures the first positive contour is three times the r.m.s. noise on the image}
\label{}
\end{figure*}

\begin{figure}
      \includegraphics[width=8cm]{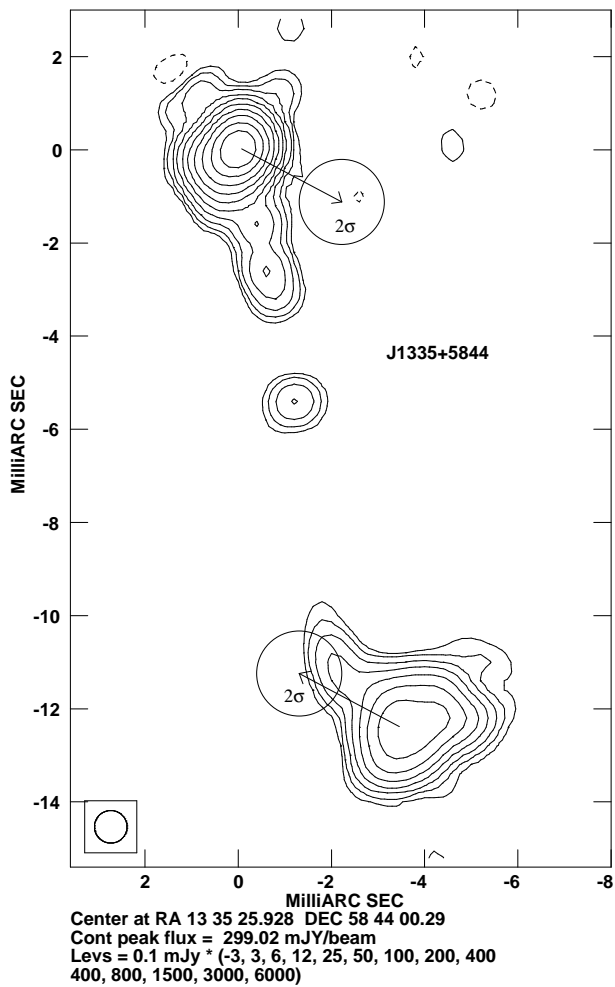}
\caption[]{J1335+5844, the arrows show the extrapolated motions in about 100 yr.}
\label{}
\end{figure}

\begin{figure*}
      \includegraphics[angle=-90,width=13.5cm]{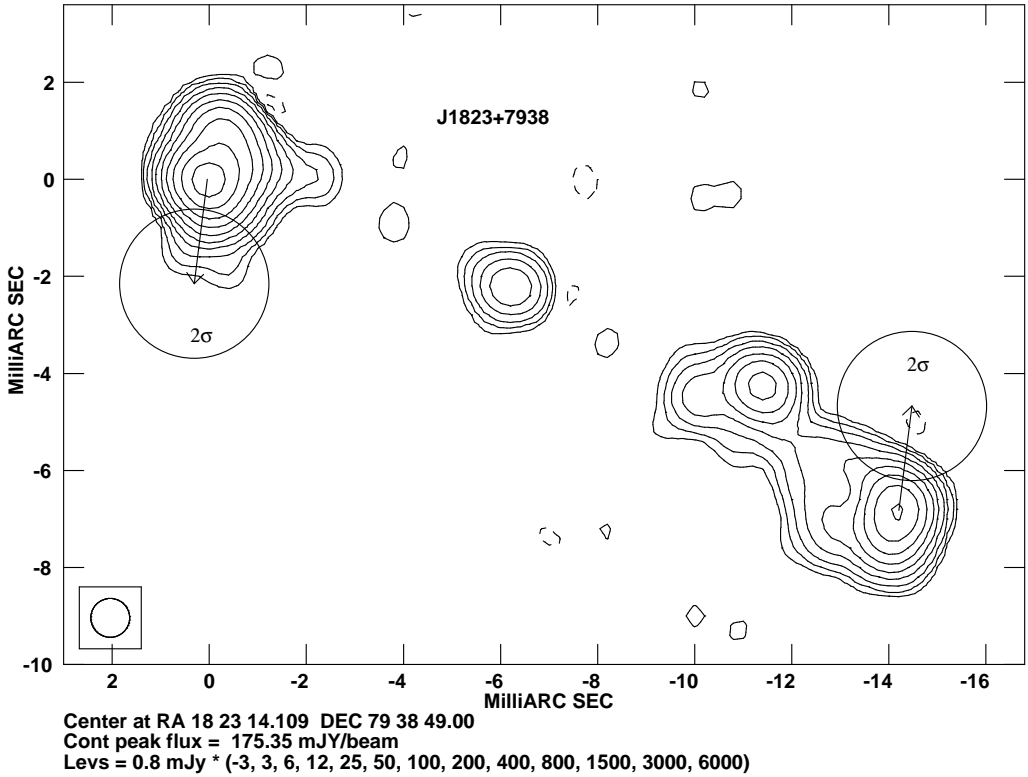}
\caption[]{J1823+7938, the arrows show the extrapolated motions in about 150 yr.}
\label{}
\end{figure*}

\section{Observations, data reduction, and results}

We observed these 3 radio galaxies with the VLBA
in two epochs for 24 hours each.
Observations for the first epoch took place in 2 observing sessions in January and March 2004,
while observations for the second epoch took place about 2.5 years later, always in two separate 
observing sessions in September and November 2006.

At each epoch each source was observed 6/8 hours at 8.4 GHz in single right polarization.

The calibration and mapping has been done in a standard way using the NRAO AIPS software. Once the final images have
been obtained, we used the DIFMAP data reduction package to model-fit the radio sources
with a few discrete pointlike or gaussian components on the fully calibrated visibilities.

The typical resolutions of the images (FWHM) are $\sim 0.8$ mas.

In J0706+4647 and J1335+5844 a very weak radio core is detected for the first time
at the center of the double structure. Such detection is truly important
to confirm the classification of the radio sorce as a Compact Symmetric Object (CSO)
and possibly to refer the hot spot motion to the center of activity.
In the present observations however, even if detected, the position of the cores 
are not well constrained, and
referring the position of the  Hot-Spot respect to the core gives too large errors.

Therefore the relative position of the 2 hot-spots have been measured in all the three objects
modelfitting the data in the UV plane with several gaussian components.
We considered as hot-spots the two strongest and more compact components derived from the modelfit.
The estimate of the accuracy and precision of this measures is usually a difficult task,
but given the high dynamical range of the images it is much better than the beam size.
In the assumption of a linear motion the errors can be estimated by the scatter of the
data themselves, which is not possible in our case with only two epochs.
In a previous work on OQ208 we could be able to estimate a 1 sigma error of 0.02 mas based on the scatter of the data, and we use the same value for the present data, which are of the same type (Stanghellini et al. in preparation).

\begin{table}[!h]
\caption{Hot-spot relative position in the 2 epochs. The position of the southern hot-spot relative
to the northern one is indicated in polar coordinates in columns 3 and 4. The difference in position
from first and second epoch is given in column 5 and 6.}
\begin{center}
\begin{tabular}{cccccc}
%\multicolumn{4}{c}{{\bf Table 2 }}\\
%\multicolumn{4}{l}{{\small VLA observations and Configurations }} \\
%\multicolumn{4}{c}{}\\
\hline
\hline
$source$      &$epoch$ & $r$ & $\theta$ & $r$ & $\theta$\\
            &       & $mas$      &$deg$& $mas$      &$deg$\\
\hline
J0706+4647&  2004& 52.126  & 109.35 &&\\
J0706+4647&  2006 &52.059 & 109.45 &0.11&-145.79\\
J1335+5844&  2004& 12.964 & -164.84 &&\\
J1335+5844&  2006 &12.891 & -165.18 &0.11&61.41\\
J1823+7938&   2004& 15.765 & -115.60 &&\\
J1823+7938&   2006& 15.745 & -115.38&0.06&-7.20 \\
\hline
\end{tabular}
\end{center}
\end{table}

We present in Tab. 2 the relative position of the two hotspots for each epoch and the difference between the two epochs. 
The errors are rather large, but there is a clear evidence of a transverse motion respect to the jet axis.
Assuming the motion of the 2 hotspots is symmetric with respect to the center of activity, the more likely explanation for
the transverse hotspot motion is precession of the jet axis. An alternative interpretation of the different position seen
in the two epochs is a change in the structure of the radio source. 
We note that the hotspots show trail of extended emission consistent with the
direction of the motion, supporting the hypothesis of real motion. 

To make the motion visible, in Fig. 1,2, and 3 we show the extrapolated motions in 500 years for the radio source J0706+4647 ($v=\sim0.4c$, assuming z=0.5), in 100 years for J1335+5844 ($v=\sim0.4c$,  photometric z=0.57), and in 150 years for J1823+7938 ($v=\sim0.15c$, z=0.224).
The arrows do not show the exact position of the components after that time, because we do not know
if the motion is constant and which is the period of the jet precession (if this is the real nature of the motion). It is only shown to make it possible to evaluate by eye the direction and the velocity of the motion. 

It seems surprising that we find these first cases of transverse motions in all the three objects 
observed , but we are not aware of any bias we could have put in our selection criteria.
These objects deserve to be observed again together with other candidates of the original list, and
it is important to monitor all CSOs of a complete sample to find how frequent tranverse motions are, but we are confident we revealed a real phenomenon.  

\section{Conclusions}

We suggest a scenario where the jet precession strongly influences the size and morphology
of a radio source. When the precession is negligible the radio source expands normally 
to become an FRI or an FRII.
When the jet precession is small the jet is able to travel through the 
channel, void of ISM particles because the transverse shocks it
created, and the hot spot has a slow transverse motion, as in the objects
presented in this work.
When the jet precession is large, the jet hits the ISM at a shorter distance
and an S shape is formed, as sometime seen in GPS radio galaxies (e.g. B0500+019, Stanghellini et al. 2001).

Rapis precession is likely to inhibit or slow down at least the radio source growth because the
head of the hotspot continuously hits new regions of the ISM. These objects
with a precessing jet may represent population of CSOs older than those
whith expansion speed measured so far, and may help to explain
the higher fraction of CSOs in samples of radiogalaxies with respect to the 
predictions of the models (Snellen et al 2000, Alexander et al. 2000).

\acknowledgements{The VLBA is operated by U.S. National Radio Astronomy Observatory which is a facility of the 
National Science Foundation, operated under cooperative agreement by Associated Universities, inc.}


\begin{thebibliography}{}
\bibitem{} Alexander, P. 2000, MNRAS 319, 8
\bibitem{} Dallacasa, D., Stanghellini, C., Centonza, M., Fanti, R.: 2000, A\&A 363, 887
\bibitem{} Dallacasa, D., Tinti, S., Fanti, C., Fanti, R., Gregorini, L., Stanghellini, C.,
Vigotti, M.: 2002a, A\&A 389, 115
\bibitem{} Dallacasa, D., Falomo, R., Stanghellini, C.: 2002b, A\&A 382, 53 
\bibitem{} Gugliucci, N.E., Taylor, G.B., Peck, A.B., Giroletti, M.: 2005, ApJ 622, 136
\bibitem{} O'Dea, C.P., Stanghellini, C., Baum, S.A., Charlot S.: 1996, ApJ 470, 806
\bibitem{} Murgia, M., Fanti, C., Fanti, R., et al.: 1999, A\&A 345, 769
\bibitem{} Polatidis,A.G., Conway, J.E.: 2003, PASA 20, 69
\bibitem{} Readhead, A.C.S., Xu, W., Pearson, T.J., Wilkinson, P.N., Polatidis, A.G.: 1994 in Compact Extragalactic Radio Sources, Proceedings of the NRAO workshop held at Socorro, New Mexico, February 11-12, 1994. Edited by J. Anton Zensus and Kenneth I. Kellermann. Green Bank, WV: National Radio Astronomy Observatory (NRAO), 1994., p.17
\bibitem{} Snellen, I.A.G., Bremer, M.N., Schilizzi, R.T., Miley, G.K., van Ojik, R.: 1996, MNRAS 279, 1294
\bibitem{} Snellen, I.A.G., Schilizzi, R. T., de Bruyn, A. G.,
 Miley, G. K., Rengelink, R.B., R\"ottgering, H. J. A., Bremer, M. N.:
1998, A\&AS 131, 435
\bibitem{} Snellen, I.A.G., Schilizzi, R.T., Miley, G.K., et al. 2000, MNRAS 319, 445
\bibitem{} Stanghellini, C., O'Dea, C.P., Dallacasa, D., Baum, S.A., Fanti, R.,
Fanti, C.: 1998, A\&AS 131, 303
\bibitem{} Stanghellini, C., Dallacasa, D., Bondi, M., Xiang L.: 2000, EVN Proceedings of the 5th european VLBI Network Symposium held at Chalmers University of Technology, Gothenburg, Sweden, June 29 - July 1, 2000, Eds.: J.E. Conway, A.G. Polatidis, R.S. Booth and Y.M. Pihlström, published Onsala Space Observatory, p. 99
\bibitem{} Stanghellini, C., Dallacasa, D., O'Dea, C.P., Baum, S.A., Fanti, R., Fanti, C.: 2001, A\&A 377, 377
\end{thebibliography}
\end{document}